\documentstyle[emulateapj,psfig,pstricks,apjfonts]{article}
\newcommand{\etal}{et~al. }
\newcommand{\eg}{e.g. }
\newcommand{\ie}{i.e. }
\newcommand{\ein}{{\sl Einstein} }
\newcommand{\msh}{G290.1--0.8}
\newcommand{\psr}{PSR~J1105-610}
\newcommand{\Hi}{\ion{H}{1}}
\newcommand{\Sii}{\ion{S}{2}}
\newcommand{\Ha}{H$\alpha$}

\begin{document}

\submitted{For Publication in The Astrophysical Journal: Accepted 28 August 
2001}
\title{An X-Ray Study of the Supernova Remnant G290.1--0.8}

\author{Patrick Slane\altaffilmark{1},
Randall K. Smith\altaffilmark{1},
John P. Hughes\altaffilmark{2},
and Robert Petre\altaffilmark{3}}

\altaffiltext{1}{Harvard-Smithsonian Center for Astrophysics, 60 Garden Street,
Cambridge, MA 02138}
\altaffiltext{2}{Department of Physics and Astronomy, Rutgers, The State
University of New Jersey, 136 Frelinghuysen Road, Piscataway, NJ 08854-8019}
\altaffiltext{3}{Laboratory for High Energy Astrophysics, Goddard Space Flight
Center, Baltimore, MD 20771}

\accepted{August 28, 2001}

\begin{abstract}
G290.1--0.8 (MSH~11--6{\it 1}A) is a supernova remnant (SNR) whose X-ray 
morphology is centrally bright.
However, unlike the class of X-ray composite SNRs whose centers are
dominated by nonthermal emission, presumably driven by a central pulsar,
we show that the X-ray emission from \msh\ is thermal in nature, placing
the remnant in an emerging class which includes such remnants as
W44, W28, 3C391, and others. The evolutionary sequence which
leads to such X-ray properties is not well understood. Here we investigate
two scenarios for such emission: evolution in a cloudy interstellar
medium, and early-stage evolution of a remnant into the radiative phase,
including the effects of thermal conduction.
We construct models for these scenarios 
in an attempt to reproduce the observed center-filled X-ray properties of
\msh, and we derive the associated age, energy, and
ambient density conditions implied by the models. We find that for reasonable
values of the explosion energy, the remnant age is of order $(1 - 2) \times
10^{4}$~yr.  This places a fairly strong constraint on any association
between \msh\ and \psr, which would require an anomalously large velocity
for the pulsar.
\end{abstract}

\keywords{ISM: individual (\msh/MSH~11--6{\it 1}A) --- supernova remnants --- 
X-rays: interstellar}

\section{INTRODUCTION}
A distinct subset of moderate age supernova remnants (SNRs) are characterized by
an X-ray morphology which is centrally brightened, in contrast to a more
limb-brightened radio profile. In some cases (\eg 
CTA~1 [Slane \etal 1997], MSH 11--6{\it 2} [Harrus, Hughes, \& Slane 1998],
G39.2-0.1 [Harrus \& Slane 1999]) recent
observations have shown that the central emission is nonthermal in nature,
presumably associated with a pulsar-driven synchrotron nebula. For others,
however, the central emission is decidedly thermal (\eg W44 [Jones, Smith
\& Angellini 1993]; 3C391 [Rho \& Petre 1996, Chen \& Slane 2001]; 
G272.2--3.2 [Harrus \etal 2001]).
A recent list of the members of this ``mixed morphology'' class of SNRs
has been compiled by Rho \& Petre (1998).
The evolutionary characteristics which have led to such
an observed profile are not well understood. One possible scenario is that
the shells in these remnants have recently become radiative, 
thus leaving only the hot interior to persist in X-rays (Smith \etal 1985, 
Harrus \etal 1997), a picture that is enhanced by the effects of 
thermal conduction
on the temperature and density distribution in the SNR (Cox \etal 1999). 
Another suggestion is that the central emission 
measure has been enhanced by the presence of cool clouds left relatively 
intact after the passage of the blast wave to slowly evaporate in the hot 
SNR interior (Cowie \& McKee 1977, McKee \& Ostriker 1977, 
White \& Long 1991).

Early X-ray observations with the \ein Observatory (Seward 1990)
established the center-filled nature of the X-ray emission from \msh.
However, the spectral characteristics of the remnant are poorly constrained
by these data, leaving open the question of whether the centrally enhanced 
X-ray emission is thermal or nonthermal in nature.
The remnant was not observed with the ROSAT PSPC.
Optical observations of the remnant show patchy emission with an enhanced
[S II]/H$\alpha$ ratio typical of SNRs (Kirshner \& Winkler 1979, Elliott
\& Malin 1979).
In the radio band (Figure 1) \msh\ is classified as a shell-type SNR 
(Green 1998) although the morphology is actually somewhat ``plateau-like'' 
with the exception of a partial shell along the western limb along with 
other patchy structures interior to the SNR boundary (Whiteoak \& Green 1996).
The remnant is $10^\prime \times 14^\prime$ in size, 
distinctly elongated in the southeast-northwest direction. This is roughly 
aligned with the Galactic plane, much like other ``bilateral'' SNRs whose
morphologies may be the result of magnetic field structures near the plane
(Gaensler 1998).
A young pulsar (J1105-6107) has 
been discovered nearby the SNR (Kaspi \etal 1997). Its location is well 
outside the remnant, along the axis of elongation (and outside the field
of view in Figure 1).  The spin-down age of the pulsar is 
$\sim 60$~kyr, considerably larger than the expected age for a remnant which is
still X-ray luminous. Moreover, as pointed out by Kaspi \etal (1997), the age
is even larger for a braking index $n < 3$, which may be more typical of
pulsars. These standard calculations assume that the current spin period $P$
is much longer than the initial period $P_0$ (i.e. that the pulsar has spun
down considerably since birth). Relaxing this constraint can lead to an
inferred pulsar age that is consistent with that of the remnant.
We discuss constraints on the association of J1105-6107 with \msh\ in 
Section 4 below.

\begin{figure*}[tb]
\pspicture(0,12)(18.5,21)

\rput[tl]{0}(0.0,20.8){\epsfxsize=9.2cm
\epsffile{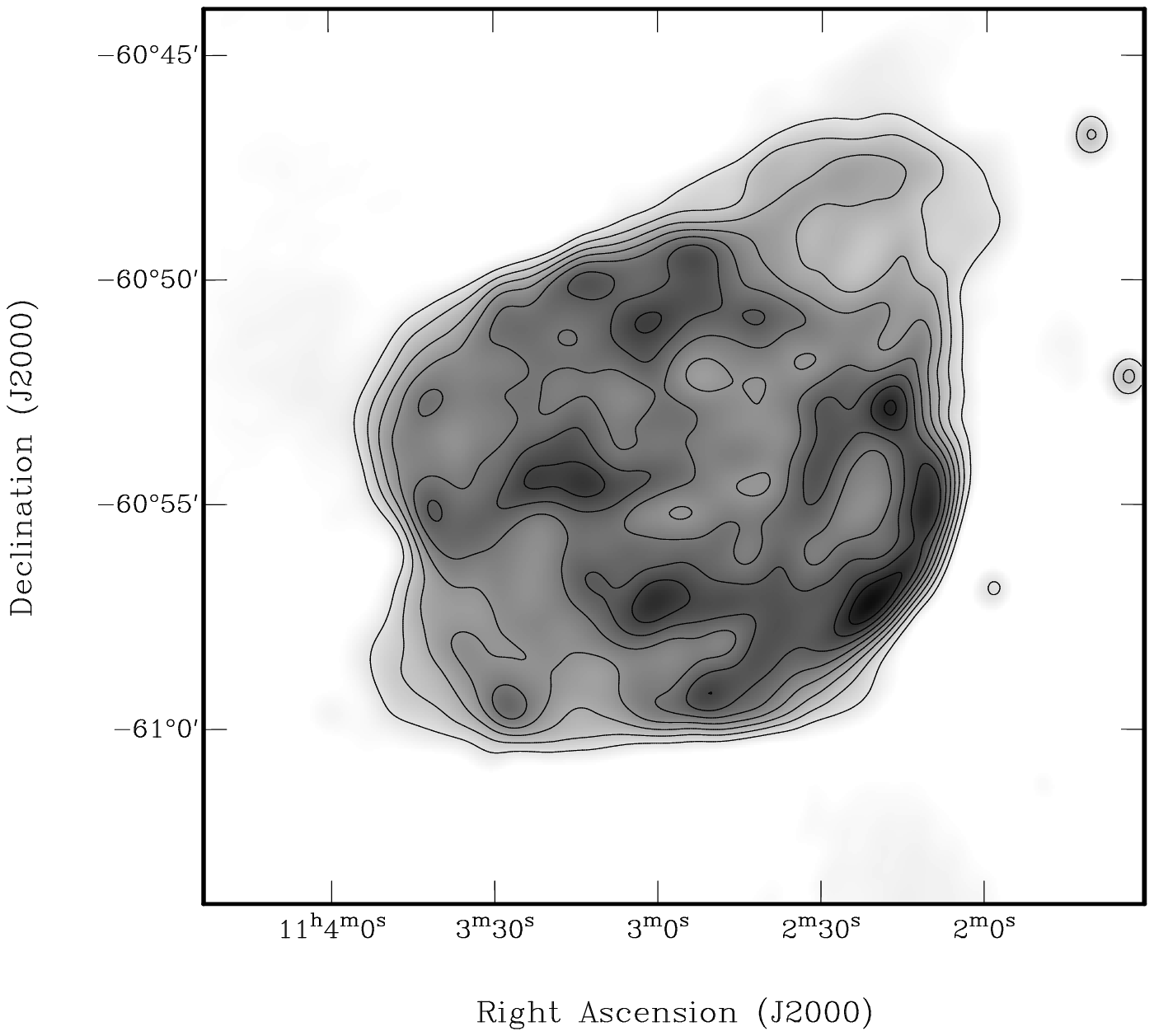}}

\rput[tl]{0}(9.75,20.7){\epsfxsize=8.0cm
\epsffile{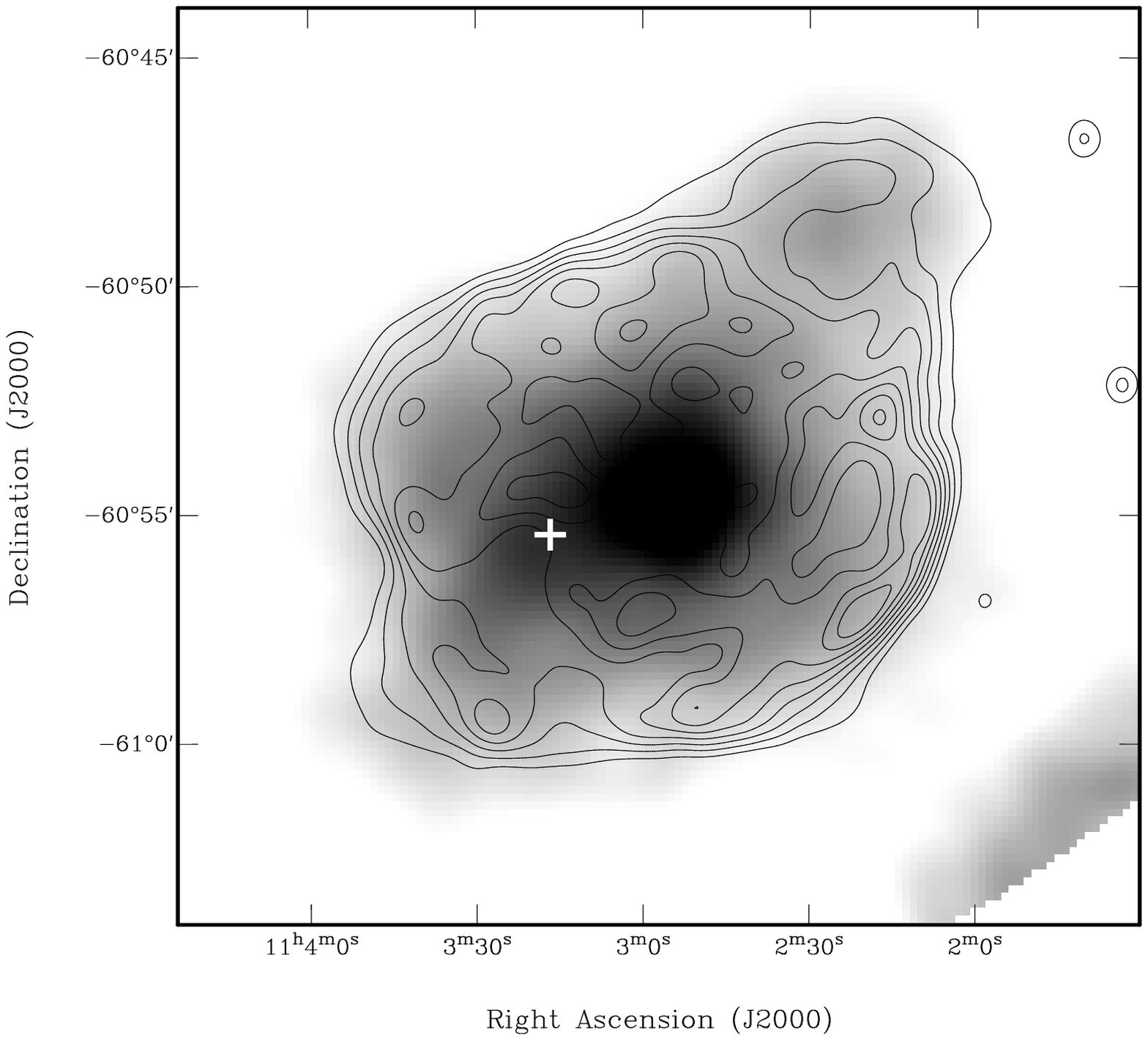}}

\rput[tl]{0}(0,12.7){
\begin{minipage}{8.75cm}
\small\parindent=3.5mm
{\sc Fig.}~1.---
Radio continuum image (843 MHz) of \msh\ from
MOST (Whiteoak \& Green 1996). Contours start at 48~mJy~beam$^{-1}$ and
increase inward in steps of 40~mJy~beam$^{-1}$.
\end{minipage}
}

\rput[tl]{0}(9.7,12.7){
\begin{minipage}{8.75cm}
\small\parindent=3.5mm
{\sc Fig.}~2.---
Background-subtracted, vignetting-corrected ASCA GIS image
(0.5-10 keV) with radio contours from MOST. The point source observed in the
ROSAT HRI is indicated by a cross.
\end{minipage}
}

\endpspicture
\end{figure*}

The distance to \msh\ is not well known. \Hi\ measurements 
(Goss \etal 1972, Dickel 1973) show absorption out to $V_{LSR} \sim -25 {\rm\ 
km\ s}^{-1}$ indicating that the remnant lies beyond the tangent point with
$D > 2.9$~kpc. Fabry-Perot measurements of the \Ha\ line (Rosado
\etal 1996) appear to show an emission component at $V_{LSR} \sim +12 {\rm\
km\ s}^{-1}$ associated with the SNR, indicating a distance $D =
6.9$~kpc. The emission is weak, however, and the \Ha\ emission is 
quite complex in this region. The \Hi\ profiles from Goss \etal (1972) indicate
absorption at positive velocities as well, but those measured by Dickel (1973)
do not. Finally, based upon a relatively low observed \Sii/\Ha\ ratio ($\sim
0.5$), Elliot \& Malin (1979) argue that \msh\ is a large, old remnant which, 
given its angular size, could be at a distance of $\sim 12-14$~kpc. 
However, studies of the \Sii/\Ha\ ratio for SNRs in M33 (Smith et al. 1993, 
Gordon et al. 1998) 
show a very large scatter in the variation of this ratio with remnant size, 
making such an argument questionable (as also noted by Rosado et al. 1996). 
For the purpose of discussion here,
we express the distance as $d_7 = (d/7 {\rm\ kpc})$ and scale all
derived quantities accordingly.

Here we report on X-ray observations of \msh\ with the Advanced Satellite for
Cosmology and Astrophysics (ASCA) in which we establish the thermal nature
of the central X-ray emission and address the observed spectral and spatial
characteristics. In Section 2 we discuss the observations
and the results of the spectral analysis. We then investigate two scenarios
to explain the X-ray characteristics: a cloudy ISM interpretation, using
the similarity solution derived by White and Long (1991); and an SNR
with thermal conduction interpretation, 
using a shock code to follow the evolution of model remnants  until properties 
representative of those observed are achieved. Finally, we present a 
discussion of these results.

\section{X-RAY OBSERVATIONS}
\msh\ was observed with the ASCA observatory for 40 ks on March 1, 1994. 
Due to observation at small Earth angle, however, standard screening
resulted in the rejection of a large fraction of the SIS data.
A repeat observation was performed on January 16, 1995. The
resulting SIS data were of somewhat reduced quality from that originally
expected because of the degradation of the SIS detectors in 4-CCD mode.

The GIS image ($0.5 - 10$~keV) of \msh\, with radio contours from the MOST 
image (Whiteoak \& Green 1996), is shown in Figure 2. Here the X-ray
data have been
smoothed with a 30 arcsec Gaussian followed by a background subtraction and 
vignetting correction. The boundary of the X-ray emission
is strikingly similar to the radio boundary,  but the morphology is centrally
bright in X-rays while the radio emission shows no evidence
of any enhancement associated with the bright X-ray center. 
The position of the peak X-ray emission is at roughly RA$_{2000}$:
11$^h$03$^m$, Dec$_{2000}$: $-60^\circ$54$^\prime$.

X-ray images of
composite SNRs (\ie those containing a central pulsar-driven synchrotron 
nebula) show distinct morphological differences above and below $\sim 1.5$~keV,
with the high energy tail of the power law spectrum from the centrally
located synchrotron nebula dominating above 1.5~keV, where the thermal
emission from the shell has fallen off
(see Harrus, Hughes, \& Slane 1997, 
Harrus \& Slane 1998).
We find no such energy dependence for the X-ray morphology of \msh; the
spectral characteristics appear relatively uniform over the remnant.

To determine the spectral properties, we have carried out joint
spectral fits using the ASCA GIS and SIS data.
For the GIS data we have used the outer 
regions of the detectors for extraction of background spectra. 
Use of blank-sky data for background, which were obtained at a high Galactic 
latitude (and thus do not include any contribution
from the Galactic ridge emission in the direction of \msh) yields similar
results for the spectral analysis. Background for the SIS was taken from
blank-sky fields. 

All GIS2 and GIS3 data were merged to produce a single spectrum. In the earlier 
observation, the GIS3 data were taken with 128 bin spectral resolution in order
to provide increased timing resolution. All GIS data were thus rebinned
to this resolution before forming the joint spectrum; no loss of intrinsic
instrument resolution is introduced in this process. A total of $\sim
55,000$
GIS events were obtained with a total effective integration time
of 139 ks. The SIS data were also merged to form a joint spectrum
yielding a total of $\sim 16,000$ events with an effective exposure of 52 ks. 

The spectra (Figure 3) are roughly described by a thermal plasma
in ionization equilibrium (Raymond \& Smith 1977) with the addition of
a weak power law component; the observed temperature is
$kT \sim 0.5$ keV with an absorption in excess of $10^{22}{\rm\ cm}^{-2}$,
and the fit yields a reduced $\chi^2$ of 1.42.
A gain adjustment of $\sim 2\%$ is required for the GIS data, a value which
is within the uncertainties of the instrument calibration.
The spectral fit is significantly improved with increased abundances for
Mg, Si, and S, yielding a final reduced $\chi^2$ of 1.02 for the spectral
values shown in Table 1.
The small hard component ($\sim 2\%$ of the unabsorbed flux) may be 
an artifact of residual Galactic
ridge emission, though we note that there is no evidence of iron-line features 
which typically characterize such emission (Yamauchi \& Koyama 1993).
Alternatively, this component could be associated with synchrotron emission
from electrons accelerated by the SNR shock. Such a component is observed
for a number of SNRs (e.g. Cas A, Tycho, Kepler, and others), including
several (SN~1006, G347.3$-$0.5, and G266.2$-$1.2) for which such emission 
completely dominates the X-ray spectrum.
In any case, the spectrum
clearly establishes that the bright central X-ray emission is thermal
in nature. Spectra taken in three 
annular regions surrounding the bright X-ray center show no strong
evidence of spectral variability, although a small variation in the
column density from region to region appears to be preferred by the
spectral fits. 

\pspicture(-0.5,12.9)(8.5,24)

\rput[tl]{0}(-0.75,24){\epsfxsize=9.0cm
\epsffile{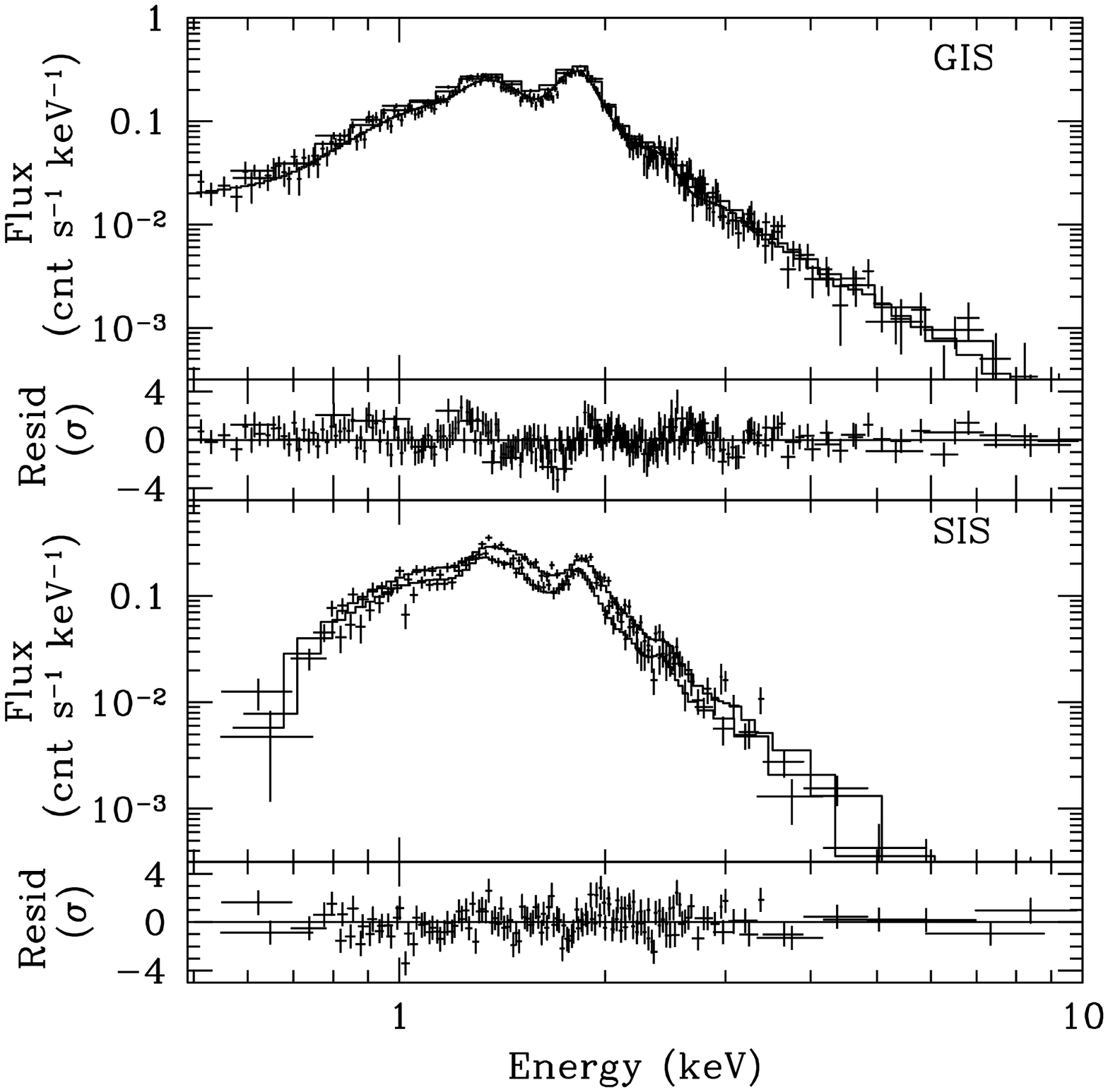}}

\rput[tl]{0}(-0.75,14.5){
\begin{minipage}{8.5cm}
\small\parindent=3.5mm
{\sc Fig.}~3.---
 X-ray spectra from \msh\ from
ASCA GIS and SIS. Histograms correspond to best-fit models as described in
the text. Residuals for the fits are shown below each spectrum.
\end{minipage}
}
\endpspicture

\msh\ was observed for $\sim 15$ ks with the ROSAT HRI on 29 January 1995. A
faint source ($R = 5.6 \times 10^{-3}{\rm\ s}^{-1}$) is seen at 
RA$_{2000}$=11$^h$03$^m$19$^s$, 
Dec$_{2000}$=$-60^\circ$55$^\prime$17$^{\prime\prime}$ which is 
within $\sim 3$~arcsec of an unidentified 14th magnitude optical source, 
consistent with the position uncertainty in the HRI.
The X-ray source is
faint and does not contribute significant distortion to the overall
brightness profile of the diffuse emission; its position is marked in
Figure 2. There is no compact source
of emission detected in the centrally enhanced region of the remnant.

\section{INTERPRETATION}
Having spectrally ruled out a central plerion scenario for \msh, 
thus placing the remnant in the same class as W44 and other ``mixed
morphology'' SNRs,
we have considered two alternative scenarios for producing the observed 
center-filled X-ray morphology: evolution in a cloudy ISM (Cowie \& McKee 
1977; White \& Long 1991), and an SNR for which thermal
conduction effects have smoothed the temperature distribution. 
Below we discuss each case and compare the model predictions with observed 
results from \msh. 

\pspicture(0,-2.0)(8.5,4)
\rput[tl]{0}(-0.5,3.5){
\begin{minipage}{8.75cm}
\small\parindent=3.5mm
\begin{center}
TABLE 1

{\sc Spectral Fit Parameters}

\vspace{1mm}

\begin{tabular}{ll} \hline
Parameter & Value$^a$ \\ \hline
$N_H$ & $(1.3 \pm 0.1) \times 10^{22}{\rm\ cm}^{-2}$ \\
$kT$ & $0.60 \pm 0.03$ keV\\
Abundance: & \\
~~Mg & $2.3 \pm 0.3$\\
~~Si & $2.4 \pm 0.3$ \\
~~S & $1.4 \pm 0.3$ \\
$F_x({\rm thermal})^b$ & $1.8 \times 10^{-10} {\rm\ erg\ cm}^{-2}
                {\rm\ s}^{-1}$ \\
$\alpha$ & $1.4 \pm 0.8$\\
$F_x({\rm nonthermal})^b$ & $2.2 \times 10^{-12} {\rm\ erg\ cm}^{-2}
                {\rm\ s}^{-1}$ \\
$\chi^2$ & 578.3 (567 degrees of freedom) \\ \hline
\end{tabular}\\
a) Uncertainties represent 90\% confidence intervals\\
b) $F_x$ = unabsorbed flux in 0.5 - 10.0 keV band~~~~~~
\end{center}

\noindent

\end{minipage}
}
\endpspicture

\subsection{Cloudy ISM Model}
The possibility that SNRs evolve in a two-phase ISM containing cold clouds
embedded in a warmer diffuse intercloud medium was first discussed by
Cowie \& McKee (1977). In this scenario, the SNR blast wave rapidly
passes the cold clouds leaving them relatively intact in the hot postshock
gas. Here, through saturated conduction, they slowly evaporate and increase
the central emission measure. White \& Long (1991) developed a similarity
solution for such evolution which incorporates two new parameters to the
standard Sedov (1959) solution, namely $C$ (the ratio of ISM mass in clouds
to that in the intercloud medium), and $\tau$ (the ratio of the cloud
evaporation timescale to the SNR age).
For appropriate values of $C$ and $\tau$, centrally bright X-ray morphologies
are possible.

To investigate the cloudy ISM scenario for \msh, we constructed models covering
a range of ($C,\tau$) values. Using
the emission measure derived from the GIS data, we varied the input shock
temperature until the mean flux-weighted temperature for the model 
matched that derived from the data. 
We then constructed the radial brightness distribution for the model
and compared it with the observed profile
The model surface brightness
distribution was calculated from the model density and temperature
distributions using emissivities for a plasma
with abundances as derived from the spectral fitting. This was then
folded through the observed interstellar absorption and
the ASCA GIS and X-ray telescope spectral response, and the model 
brightness and temperature profiles were then compared with those derived
from the GIS data. As illustrated in Figure 4, an adequate description
of the data is obtained for ($\tau, C, kT_s$) values 
between (10,50,0.49~keV) and (40,150,0.51~keV). 
While the brightness distribution differs somewhat from that observed
(see further discussion in Section 4), the overall agreement is quite good for
a spherically symmetric, homogeneous model; deviations of similar magnitude are
typically encountered when attempting to fit profiles for shell-type remnants
with ideal Sedov models.
We note that the radio boundary indicated in Figure 4 is that of the most
spherical component of \msh, and this is the valued used for the remnant
radius in the (spherically symmetric) models.
The apparent extension of the X-ray boundary
beyond the radio shell is due primarily to the contribution from the extended
regions of emission in the NW and SE (although the broad point spread
of the telescope also smoothes the outer X-ray boundary).

\begin{figure*}[tb]
\pspicture(0,10.0)(18.5,21)

\rput[tl]{0}(0.0,20.8){\epsfxsize=8.8cm
\epsffile{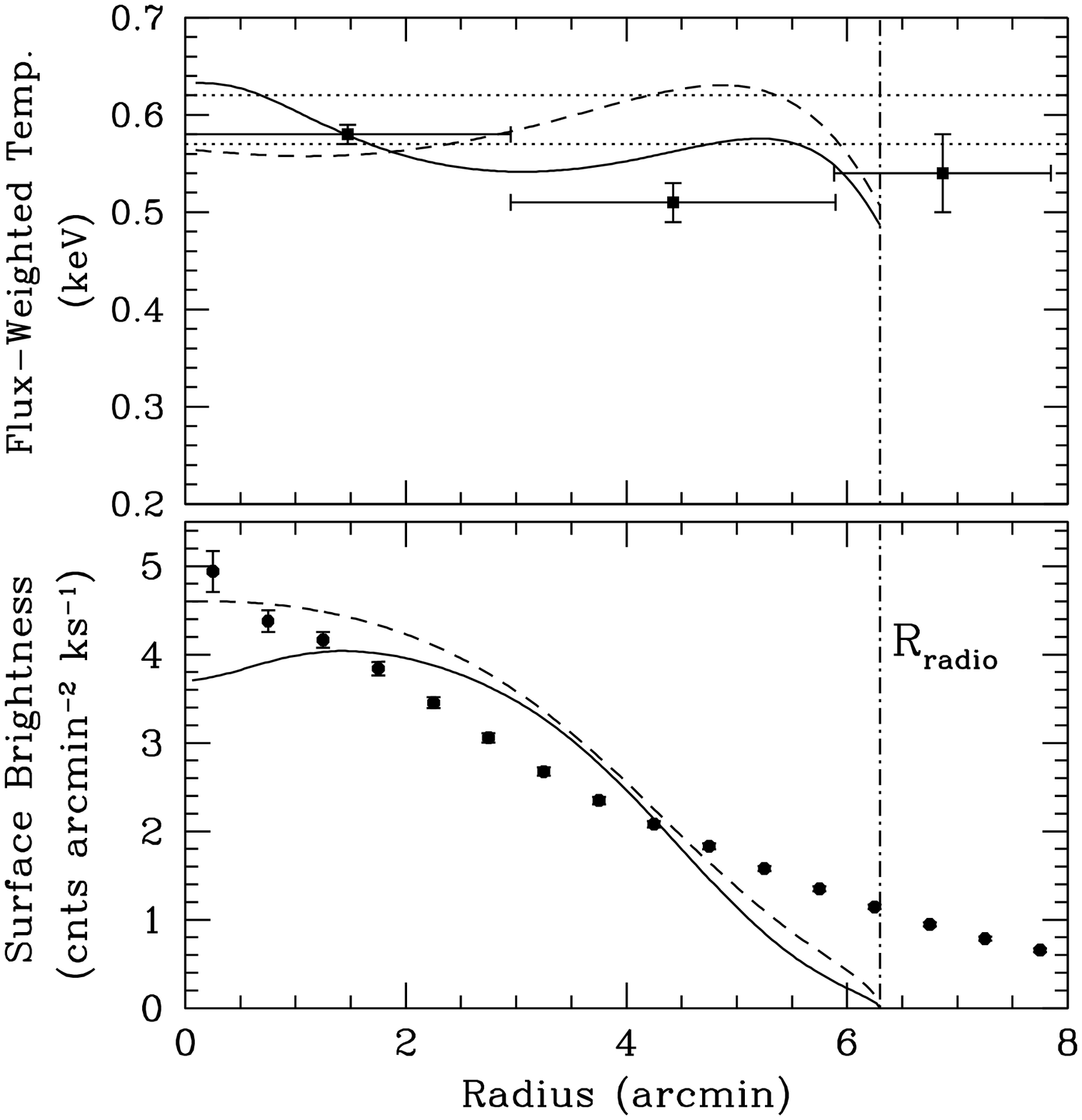}}

\rput[tl]{0}(9.55,20.7){\epsfxsize=8.8cm
\epsffile{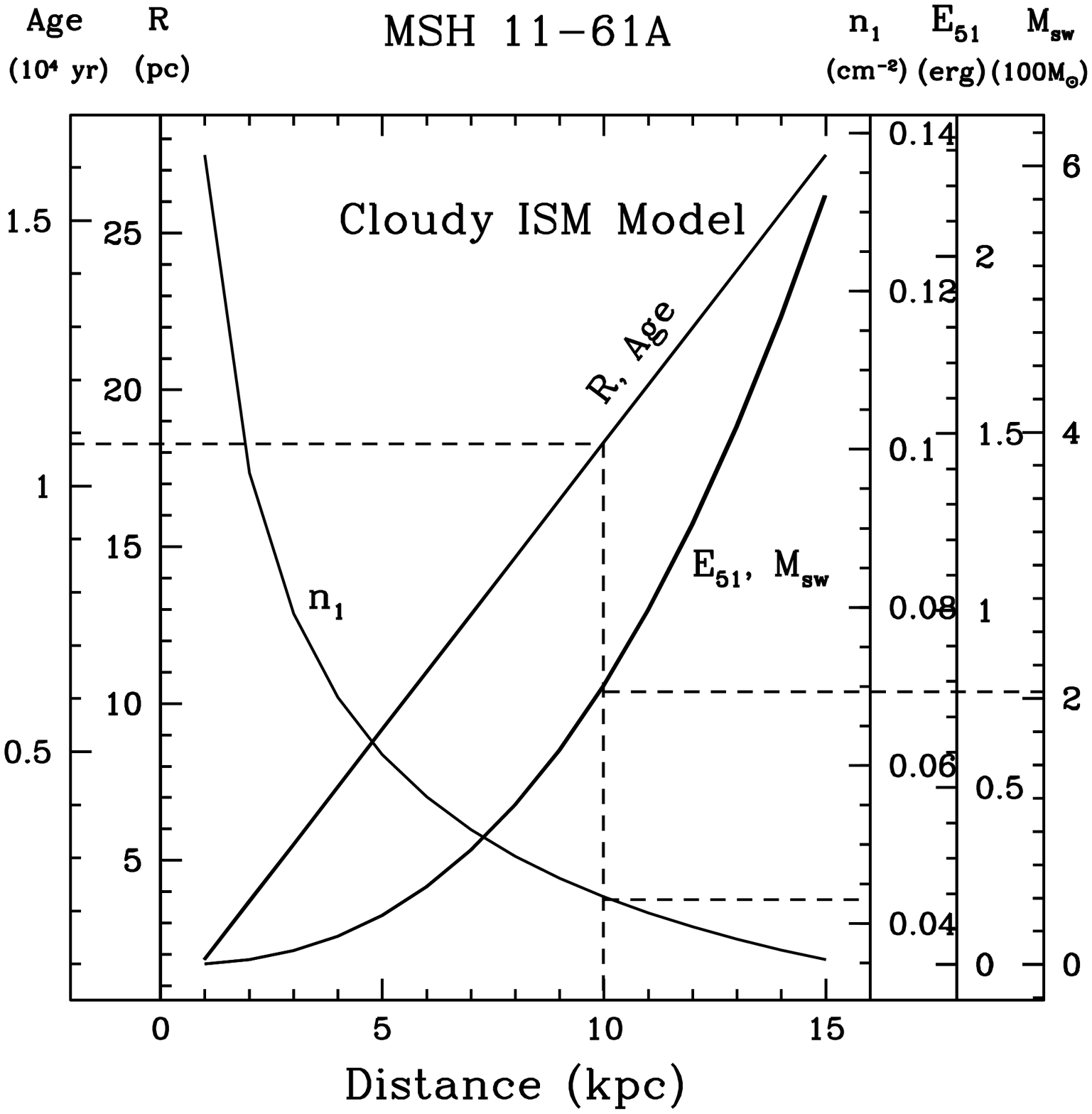}}

\rput[tl]{0}(0,11.7){
\begin{minipage}{8.75cm}
\small\parindent=3.5mm
{\sc Fig.}~4.---
Comparison of cloudy ISM models with observed X-ray surface
brightness and temperature profiles for \msh. Points correspond to data
from ASCA GIS observations. Curves
correspond to models using the similarity solutions of White \& Long (1991)
for input parameters $(\tau,C,kT_s)$ = (10,50,0.49~keV).
[solid] and (40,150,0.51~keV) [dashed].
\end{minipage}
}

\rput[tl]{0}(9.7,11.7){
\begin{minipage}{8.75cm}
\small\parindent=3.5mm
{\sc Fig.}~5.---
Derived parameters for \msh\ based upon cloudy ISM model.
Curves for radius and age, $E_{51}$ and $M_{\rm\ sw}$, and $n_1$ are
indicated with associated scales at left and right. Dashed lines indicate
example values based upon a 10 kpc distance.
\end{minipage}
}

\endpspicture
\end{figure*}

In Figure 5 we have plotted the derived properties of the remnant 
for ($\tau, C, kT_s$) values of (10,50,0.49~keV).
The dependence on assumed distance is shown explicitly. The derived value for
the explosion energy ($\sim 3 \times 10^{50}{\rm\ erg}$ is somewhat
low for the nominal $\sim 7$ kpc distance;
more reasonable values ($\sim 8 \times 10^{50}{\rm\ erg}$)
are obtained for a distance closer to $\sim 10$ kpc,
although we note that inferred explosion energies are often low with the 
W\&L model (e.g. Harrus et al. 1997).
Using the ($\tau, C, kT_s$) values of (40,150,0.51~keV) leads to a $\sim 50\%$
decrease in $n_1$, but little change in the other derived quantities.

While the cloudy ISM model reproduces the observed X-ray characteristics
of \msh\ quite well, an area of concern is the very long evaporation timescales
(10 to 40 times the age of the remnant) required by the model. 
Models for evaporation of cold, dense clouds via
saturated conduction predict relatively long evaporation timescales
(McKee \& Ostriker 1977). However, hydrodynamic simulations
of shock/cloud interactions indicate that the clouds undergo considerable
disruption (Stone \& Norman 1992); models which require that 
cloudlets remain intact on timescales of many tens of thousands of years 
(as suggested by the values if $\tau$) appear problematic. Magnetic fields 
may act to prolong the cloud 
destruction timescale (Mac Low et al. 1994) although
this may also act to reduce the evaporation process because thermal
conduction is heavily suppressed across magnetic field lines.

\subsection{Thermal Conduction Model}

In the later phases of SNR evolution, the shock front slows and
becomes radiative, as the shocked material cools immediately after
being heated.  This occurs when the shock temperature falls below
$\sim 0.1$\,keV, which, using the Sedov (1959) solution, corresponds
to an age $t \sim 2\times10^4 ({{E_{51}}\over{n_0}})^{1/3}$\,yr, with
a radius $R \sim 15 ({{E_{51}}\over{n_0}})^{1/3}$\,pc.  After this
stage, the X-ray emission above $\sim 1$\, keV from the outer shell of
the remnant effectively stops.  In addition, the soft X-ray/EUV
radiation from the radiative shock is readily absorbed, so the central
hot region of the remnant dominates the X-ray emission.

In a pure Sedov solution, however, the temperature in the center of
the remnant is extremely high, and the density low, so the total X-ray
emissivity of the center will be very low as well.  Reducing the
central temperature and increasing the density would dramatically
increase the emission.  In the White \& Long (1991) model, this
moderation was achieved by assuming evaporating clouds exist inside
the hot remnant.  Another method is to include thermal conduction, via
Coloumb collisions between the electrons and ions (Spitzer 1956),
inside the hot plasma.  This approach has been used to model the SNR
W44 (Cox \etal 1999; Shelton \etal 1999), and to make models of the
Local Bubble (Smith \& Cox 2001).  

We used a one-dimensional spherically symmetric Lagrangian shock code
({\sc odin}) (Smith \& Cox 2001) to model the temperature and density
evolution of the SNRs into the radiative phase.  {\sc odin}\
calculates the non-equilibrium radiative cooling (using the 1993
version of the Raymond \& Smith (1977) plasma emission model) from the
hot plasma at each timestep.  The electron and ion temperatures are
assumed to be in equilibrium throughout.  The magnetic field is
assumed to be frozen in the plasma, and so contributes only to the
pressure as a term $\propto n^2$.  The thermal conduction model is
based on Spitzer (1956), including the saturation limit from Cowie
\& McKee (1977).  The code has been tested against the analytic Sedov
model as well as the shock tube model (Smith 1996).

\begin{figure*}[tb]
\pspicture(0,10.0)(18.5,21)

\rput[tl]{0}(0.0,20.8){\epsfxsize=8.8cm
\epsffile{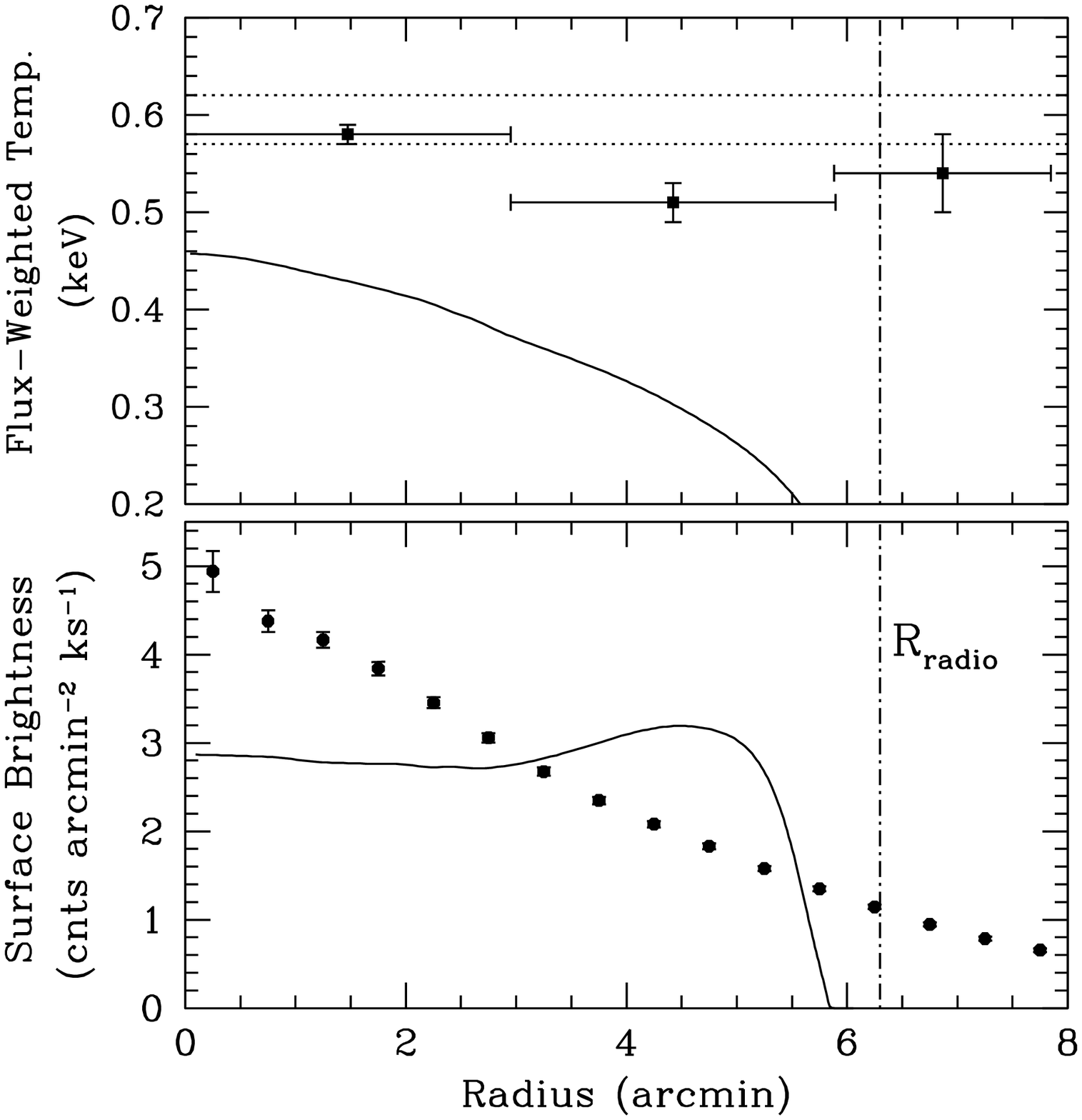}}

\rput[tl]{0}(9.95,20.7){\epsfxsize=8.8cm
\epsffile{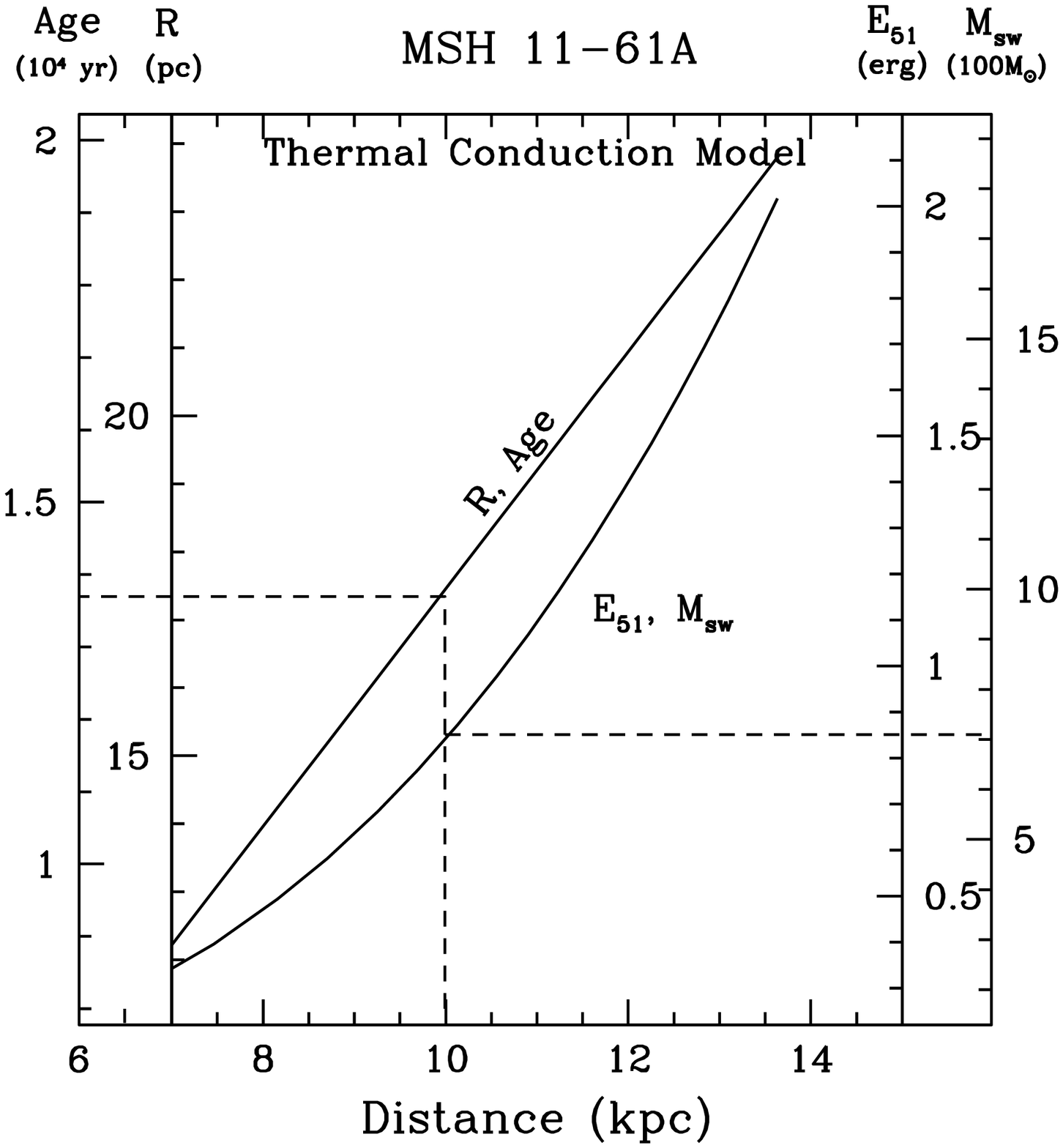}}

\rput[tl]{0}(0,11.7){
\begin{minipage}{8.75cm}
\small\parindent=3.5mm
{\sc Fig.}~6.---
Comparison of thermal conduction model with observed
X-ray surface brightness and temperature profiles for \msh.
Points correspond to data from ASCA GIS observations. Curves
correspond to the thermal conduction model which best reproduces the average
surface brightness and mean temperature of the remnant.
\end{minipage}
}

\rput[tl]{0}(9.7,11.7){
\begin{minipage}{8.75cm}
\small\parindent=3.5mm
{\sc Fig.}~7.---
Derived parameters for \msh\ based upon radiative model
with thermal conduction. Curves for radius, age, $E_{51}$ and $M_{\rm\ sw}$
are indicated with associated scales at left and right. Dashed lines indicate
example values based upon a 10 kpc distance.
\end{minipage}
}

\endpspicture
\end{figure*}

We constructed models for a range of ($E, n_0$) values and followed
these to 50,000 years, by which time all had become radiative.  We
then calculated the NEI X-ray spectrum for each ($E, n_0, t$) point,
folded this through the ASCA GIS response, and found the predicted
surface brightness.  We restricted our search to those models that
matched the observed surface brightness (0.0025 GIS
counts~sec$^{-1}$~arcmin$^{-2}$). The temperature profiles of these models are
all relatively flat compared to models without thermal conduction
({\it e.g.}, Sedov models).  The central temperature is a function of
ISM density $n_0$\ but largely independent of supernova energy $E$, given
the constraint on the surface brightness.

Since the observed temperature profile is quite flat at $\sim
0.6$\,keV, we were able to eliminate models with $n_0 \ge
2.0$\,cm$^{-3}$\ because they were too cool throughout.  Conversely,
models with $n_0 \le 0.5$\,cm$^{-3}$\ were too
limb-brightened.  We were therefore driven to models with $n_0 =
1$\,cm$^{-3}$.  In Figure 6 we show the flux-weighted temperature
profile and the surface brightness for the model with $E_{51} = 1.0$\
and $n_0 = 1.0$\,cm$^{-3}$.  Clearly the flux-weighted
temperature is systematically too low, although we note that
the derived temperature is dependent upon the elemental abundances as well as
the state of ionization equilibrium. We have modeled these parameters as well,
but the important line features that are sensitive to the parameters are
poorly constrained in broad-band fits to low spectral resolution data. 
The model surface brightness is also limb-brightened, in contrast to the
centrally-peaked remnant morphology.
Although the center of the remnant in this model has $kT > 0.7$\,keV, the
emission from the edge dominates because the shock has not yet become
radiative. It is important to note, however, that the limb brightness in this
model is only $\sim 10\%$ higher than that for the remnant center, a condition
far from that for a standard Sedov phase remnant. In Figure 7 we show the model
results for the age, size, and explosion energy as a function of the
assumed distance, for an ISM density of 1 cm$^{-3}$.  Our models do
predict the remnant age and radius as a function of explosion energy,
but as the data do not constrain this parameter, we can report only
ranges.  For explosion energies $E_{51} = 0.3, 1.0, 2.0$, the remnant
ages are 10, 16, and 22 kyr, and the radii 6.2, 9.6, and 12 pc,
respectively.

\section{DISCUSSION}

Both of the models considered here yield X-ray emission that is centrally
enhanced relative to that expected from a pure Sedov model. 
The observed brightness profile is actually steeper in the center than either
model predicts, however, a situation that holds for similar modeling
of W44 as well (Harrus et al. 1997, Shelton et al. 1999). One possible scenario
to explain this could be the presence of ejecta in the central regions.
The X-ray spectrum of \msh\ does show enhanced abundances of Mg, Si, and 
possibly Si, which could be consistent with this interpretation. However, 
there is currently no evidence that the measured abundance varies with 
radius, which would be required in such a scenario. 
Observations with higher angular resolution are of considerable
interest in order to permit a more sensitive search for such spectral
variations. We note that a second scenario which could modify the
brightness profile in such a way as to boost the central emission
might be evolution into a stellar wind density distribution. This
would consist of a pre-existing density profile which declines rapidly
with radius, thus yielding relatively more mass in the central regions.
Incorporation of such profiles into the models discussed here is beyond
the scope of this paper.

As illustrated in Figures 5 and 7, the derived properties for
\msh\ differ for the two distinct models. However, the basic conclusion
supported by each is that, for reasonable values of the explosion
energy ($E_{51} \approx 0.5 - 1$), the remnant age is of order
$(1 - 2) \times 10^4$~yr, the distance is $\sim 8 - 12$~kpc, and
the amount of swept-up mass is very large. \msh\ is clearly in the
mid-to-late phase of evolution.
Our thermal conduction models do not predict that the shock in \msh\ has gone
radiative, unlike similar models used successfully for W44 (Shelton et al.
1999). This explains the slightly limb-brightened brightness profile for the
model illustrated in Figure 6; it appears that thermal conduction models
for mixed morphology SNRs only become centrally dominated in X-rays once the
shock has become radiative. 
In the late phase of evolution, after the shock has become radiative,
we expect the formation of a dense,
thin outer shell that should be observable in HI. Such a shell is observed
in W44 (Koo \& Heiles 1995), for example, and HI observations of 
\msh\ are thus of considerable importance. We note that our thermal conduction
models for \msh\ are unable to reproduce even the average surface brightness
and mean temperature for cases where the shock is radiative. If HI observations
do reveal a dense outer shell, this will imply that other parameters for the
X-ray emitting plasma need modification if a thermal conduction model is to
hold.

The X-ray characteristics of \msh\ can also be used to assess the likelihood
of an association between the remnant and \psr. The component of the
pulsar velocity in the plane of the sky is given by $v_p = \beta D/t$ where
$D$ is the distance to the system, $t$ is the age, and $\beta$ is the angular
separation (in radians) between the pulsar and the SNR center. From 
pulsar population
studies, velocities are distributed in two roughly Gaussian components with
characteristic speeds of $175 {\rm\ km\ s}^{-1}$ and 
$700 {\rm\ km\ s}^{-1}$, representing 86\% and 14\% of the population,
respectively (Cordes \& Chernoff 1998).

For a remnant in the Sedov phase of evolution (including the case with
evaporating clouds), the shock radius can be written
$$ R_s = \left[\frac{25(\gamma + 1)^2}{8 (\gamma - 1)}
\frac{kT_s}{\mu m_H}\right]^{1/2} t$$
where $T_s$ is the shock temperature, $\mu \approx 0.6$ for cosmic abundances,
and $\gamma = 5/3$ for an ideal gas. Since $R_s = \theta D$, where $\theta$
is the angular size of the remnant, the pulsar velocity can be written
$$ v_p = 6.8 \times 10^2 \left(\frac{\beta}{\theta}\right) T_6^{1/2} 
{\rm\ km\ s}^{-1}$$
where $T_6$ is the temperature in units of $10^6$~K. For \msh\ we have
$\theta \sim 6.3$~arcmin, and \psr\ is located $\beta \sim 22$~arcmin from the
remnant center. For the cloudy ISM model described above, the inferred
shock temperature is $T_s = 4.1 \times 10^6$~K. This leads to a pulsar
velocity $v_p \approx 5.3 \times 10^3 {\rm\ km\ s}^{-1}$ which is much
larger than expected for a true association between the two objects.

For the thermal conduction model, for which the Sedov solution no
longer applies, the shock temperature is lower. Scaling the size/age relation
from Figure 7 implies a pulsar velocity $v_p \approx 4.5 \times 10^3 {\rm\ km\
s}^{-1}$, still very large when compared with other pulsars.
We conclude that \msh\ and \psr\ are unlikely to have been formed in the
same supernova explosion.

\section{CONCLUSIONS}

The ASCA observations of \msh\ establish this remnant as another member
of the ``thermal composite'' class whose X-ray morphology is dominated
by thermal emission from the center. We have investigated two models
in an effort to reproduce the observed temperature and brightness
profiles. The cloudy ISM model based on the similarity solution by
White \& Long (1991) provides a reasonable description of these
observed characteristics, and leads to a remnant of moderate age at
a distance of $\sim 8 - 11$~kpc for explosion energies in the range
$E_{51} = 0.5 - 1$. The required density of the intercloud medium is 
$\sim 0.4 - 0.05 {\rm\ cm}^{-3}$, and the cloud to intercloud mass ratio is
of order $50 - 150$. The evaporation timescale for the clouds is
$10 - 40$ times the age of the SNR, which may be problematic given that
cloud-crushing timescales are typically much shorter than this.
The observed brightness profile is more centrally peaked than the model
predicts, which may indicate a need to consider a density profile modified by a
precursor wind.

A hydrodynamic model which follows the remnant evolution toward the radiative
phase, and incorporates the effects of thermal conduction, is also
capable of increasing the central X-ray brightness and producing a somewhat
flattened temperature distribution, but the brightness and temperature profiles
differ considerably from those observed. It would appear that this model
requires modifications to the ambient medium, and perhaps the ejecta component,
in order to more closely reproduce the observed properties.
The density required to reproduce the
size and flux is $n_0 \sim 1 {\rm\ cm}^{-3}$, and
the distance range associated with the explosion
energy range used above is also $\sim 8 - 11$~kpc. 

With either of the above interpretations, the angular separation between
\msh\ and \psr\ relative to the remnant radius implies a pulsar velocity
much larger than that typical of even high velocity pulsars. This
would suggest that \psr\ is not likely to be associated with the SNR
if either of these interpretations for the evolution are correct.

\acknowledgments

This work was supported in part by the National Aeronautics and
Space Administration through grants NAG5-3486 and NAG5-2638, and
contract NAS8-39073. The authors would like to thank Ilana Harrus for 
many useful discussions on the nature of thermal composite SNRs.


\begin{references}

\reference{}
Chen, Y. \& Slane, P. 2001, ApJ - accepted for publication

\reference{}
Cordes, J. M. \& Chernoff, D. F. 1998, ApJ, 505, 315

\reference{}
Cowie, L. L. \& McKee, C. F. 1977, ApJ, 211, 135

\reference{}
Cox,~D.~P. \etal  1999, ApJ, 524, 179

\reference{}
Dickel, J.R. 1973, Astrophys. Lett., 15, 61

\reference{}
Elliot, K.H, \& Malin, D.F. 1979, MNRAS, 186, 45

\reference{}
Gaensler, B. M. 1998, ApJ, 493, 781

\reference{}
Gordon, S. M., Kirshner, R. P., Long, K. S., Blair, W. P., Duric, N., 
\& Smith, R. C. 1998, ApJS, 117, 89

\reference{}
Goss, W.M., Radhakrishnan, V., Brooks, J.W., \& Murray, J.D. 1972, ApJS, 203,
123

\reference{}
Green D.A., 1998, `A Catalogue of Galactic Supernova Remnants (1998 September
version)', Mullard Radio Astronomy Observatory, Cambridge, United
Kingdom (available on the World-Wide-Web at
http://www.mrao.cam.ac.uk/surveys/snrs/) 

\reference{}
Harrus, I. M., Hughes, J. P., Singh, K. P., Koyama, K., and Asaoka, I. 1997, 
ApJ, 488, 781

\reference{}
Harrus, I. M., Hughes, J. P., \& Slane, P. O. 1998, ApJ, 499, 273

\reference{}
Harrus, I. M. \& Slane, P. O. 1999, ApJ, 516, 811

\reference{}
Harrus, I. M. Slane, P. O., Smith, R. K., \& Hughes, J. P. 2001, ApJ, 552, 614

\reference{}
Jones, L. R., Smith, A., \& Angellini, L. 1993, MNRAS, 265, 631

\reference{}
Kaspi, V. M., Bailes, M., Manchester, R. N., Stappers, B. W., Sandhu, J. S.,
Navarro, J., \& D'Amico, N. 1997, ApJ, 485, 820

\reference{}
Kirshner, R. P. \&  Winkler, P. F., Jr. 1979, ApJ, 227, 853

\reference{}
Koo, B.-C \& Heiles, C. 1995, ApJ, 442, 679

\reference{}
Mac Low, M-M., McKee, C. F., Klein, R. I., Stone, J. M. \& Norman, M. L.
1994, ApJ, 433, 757

\reference{}
McKee, C. F. \& Ostriker, J. P. 1977, ApJ, 218, 148

\reference{}
Raymond, J. C. \& Smith, B. 1977, ApJS, 35, 419

\reference{}
Rho, J.-H., \& Petre, R. 1996, ApJ, 467, 698

\reference{}
Rho, J.-H., \& Petre, R.  1998, ApJ, 503L, 167

\reference{}
Rosado, M., Ambrocio-Cruz, P., Le Coarer, E., \& Marcelin, M. 1996, A\&A, 315,
243

\reference{}
Sedov, L. 1959, Similarity and Dimensional Methods in Mechanics (New
York:Academic)

\reference{}
Seward, F. D. 1990, ApJS, 73, 781

\reference{}
Shelton,~R.~L. \etal 1999, ApJ, 524m 192

\reference{}
Slane, P., Seward, F. D., Bandiera, R., Torii, K., \& Tsunemi, H. 1997,
ApJ, 485, 221

\reference{}
Smith, A., Jones, L. R., Watson, M. G., Willingale, R., Wood, N., \& Seward, F.
D. 1985, MNRAS, 217, 99

\reference{}
Smith~R.~K. 1996 - Ph.D. thesis, Univesity of Wisconsin

\reference{}
Smith~R.~K. \& Cox,~D.~P. 2001, ApJS, 134, 283

\reference{}
Smith, R. C., Kirshner, R. P., Blair, W. P., Long, K. S., \& Winkler, P. F. 
1993, ApJ, 407, 564

\reference{}
Spitzer, L. 1956, ``Physics of Fully Ionized 
Gases'', (New York: Interscience Publishers), p. 87

\reference{}
Stone, J. M. \& Norman, M. L. 1992, ApJ, 390, L17

\reference{}
Vasisht, G., Aoki, T., Dotani, T., Kulkarni, S. R., Nagase, F. 1996, ApJ,
456, L59

\reference{}
White, R. L. \& Long, K. S. 1991, ApJ, 373, 543

\reference{}
Whiteoak, J. B. Z. \& Green, A. J. 1996, A\&AS, 118, 329

\reference{}
Yamauchi, S. \& Koyama, K. 1993, ApJ, 404, 620

\end{references}
\end{document}